**Title**: Thermally Driven Analog of the Barkhausen Effect at the Metal-Insulator Transition in Vanadium Dioxide

**Authors**: Benjamin Huber-Rodriguez[1], Siu Yi Kwang[2], William J. Hardy[3], Heng Ji[1], Chih-Wei Chen[1], Emilia Morosan[1,4,5], Douglas Natelson[1,5,6]

**Affiliations**:
1. Department of Physics and Astronomy, Rice University, Houston, TX 77005
2. Department of Physics, National University of Singapore, Singapore 117551
3. Applied Physics Program, Rice Quantum Institute, Rice University, Houston, TX 77005
4. Department of Chemistry, Rice University, Houston, TX 77005
5. Department of Materials Science and Nanoengineering, Rice University, Houston, TX 77005
6. Department of Electrical and Computer Engineering, Rice University, Houston, TX 77005

**Abstract**:

The physics of the metal-insulator transition (MIT) in vanadium dioxide remains a subject of intense interest. Because of the complicating effects of elastic strain on the phase transition, there is interest in comparatively strain-free means of examining $VO_2$ material properties. We report contact-free, low-strain studies of the MIT through an inductive bridge approach sensitive to the magnetic response of $VO_2$ powder. Rather than observing the expected step-like change in susceptibility at the transition, we argue that the measured response is dominated by an analog of the Barkhausen effect, due to the extremely sharp jump in the magnetic response of each grain as a function of time as the material is cycled across the phase boundary. This effect suggests that future measurements could access the dynamics of this and similar phase transitions.

**Text**:

Vanadium dioxide is a showcase for the rich physics possible in correlated transition metal oxides, where strong electron-electron and electron-phonon interactions lead to competition between phases with vastly differing electronic properties. With zero strain, $VO_2$ undergoes a first-order transition at 65º C between a high temperature, rutile, metallic phase (R) and a low temperature, monoclinic, insulating phase (M1)[1,2]. Both phases are paramagnetic, with the M1 phase having a roughly temperature-independent susceptibility $\chi' \approx 0.95 \times 10^{-6}$ emu/g, and the R phase having a temperature-dependent susceptibility (direction-averaged) starting at $\chi' \approx 8.1 \times 10^{-6}$ emu/g at the transition and falling with increasing temperature [3,4]. The higher temperature response is attributed to spin paramagnetism of comparatively low mobility vanadium $d$ electrons in the narrow band cutting the Fermi level in the metallic state [4]. The detailed role of electronic correlations in both the transition and the properties of the metallic state [5] remain a subject of debate.

Studies of the transition are particularly challenging because of the strong effects of strain on the phase boundaries, a perturbation that has been examined using single-crystal nanobeams[1,6,7]. Strain has a similarly strong impact on the doping of $VO_2$ with interstitial hydrogen and the resulting material response[8]. Even the presence of lithographically defined contacts for electronic transport characterization can act as a perturbative source of strain. Thus, there is a strong need for a non-contact means to assess the physical properties of $VO_2$ across the metal-insulator phase boundary.

Inductive studies of $VO_2$ powder offer one potential route toward non-contact assessment of the material's electronic and magnetic properties. In this Letter we report inductive bridge measurements intended to compare the response of an inductor filled with nominally strain-free $VO_2$ powder with that of an empty but otherwise identical inductor. As the temperature of the coils is swept across the transition, rather than observing a step-function increase in inductive response proportional to the susceptibility, $\chi'(T)$, we instead find a large signal proportional to the time derivative, $(d\chi'/dT)(dT/dt)$. We ascribe this signal to an analog of the Barkhausen effect[9], detectable due to the abrupt transition of individual crystals within the powder. This effect may provide an alternative means of examining the dynamics of this and other metal-insulator transitions in contact-free powders.

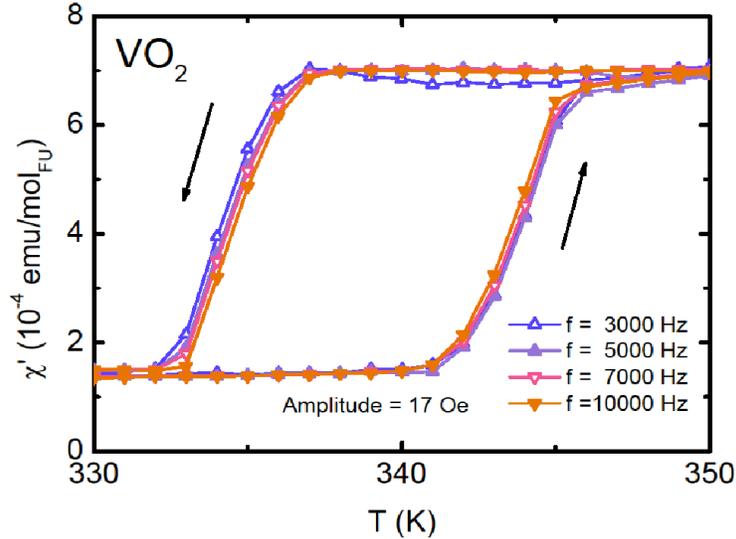

**Figure 1**: AC magnetic susceptibility of the $VO_2$ powder measured with a commercial magnetometer upon warming. Data at each temperature point were acquired with various AC frequencies after allowing the sample to thermally equilibrate at each point.

The starting material was vanadium (IV) oxide at 99% nominal purity and packed under argon (Alfa Aesar). To eliminate any trace impurity of $V_2O_5$, we baked this material under 5 mB argon gas, ramping up to $900°$ C over thirty minutes, remaining at this temperature for thirty minutes, then cooling to room temperature. Measurements of the AC magnetic susceptibility (ACMS) performed on the $VO_2$ powder using the ACMS option in a commercial magnetometer (Quantum Design) show the expected step function in $\chi'$ at the transition, rounded by ensemble averaging over the individual crystals in the powder. As shown in Fig. 1, the measured $\chi'$ is consistent with literature values[3] and shows the characteristic thermal hysteresis expected of such a first-order phase transition. In the ACMS measurements, the data are acquired after allowing the sample to equilibrate at each temperature. The measured $\chi'$ is essentially independent of frequency over the acoustic frequency band, and the dissipative component $\chi''$ (not shown) is featureless across the transition within the resolution of the measurement.

Measurements were taken using an AC inductance bridge, shown in Figure 2a. A lock-in amplifier supplied the AC excitation via a ratio transformer to two identical inductors, one containing the prepared $VO_2$ powder. The inductors were nominally identical 200 turn coils, 8.5 cm in length, wound with copper wire around hollow fused silica tubes (4.0 mm outer diameter, 2.25 mm inner diameter). The filled inductor held 0.95 grams of $VO_2$ powder. The lock-in amplifier performs a phase-sensitive measurement of the difference in voltage across the two inductors. The bridge was initially balanced at room temperature to account for any minor geometric variation between the coils. Assuming otherwise identical coils and uniform heating, conventionally the inductive phase of the bridge measurement is expected to show an off-balance signal proportional to changes in $\chi'$ as the temperature is varied, while the dissipative phase measurement of the off-balance signal should show a contribution from changes in $\chi''$.

To maximize temperature uniformity during the measurement, the inductors were placed in cylindrical cavities hollowed out of a solid block of aluminum (76.2 mm x 50.8 mm x 19.6 mm) and heat sunk via thermal grease. The block geometry relative to the coil lengths was chosen to minimize interactions between the inductor fringing fields and the aluminum. The aluminum block was heated uniformly and at a controllable rate via heating of a nichrome wire wound back and forth through holes drilled in the block. The cooling rate is limited by thermal coupling of the Al block to the ambient environment. A temperature sensor was placed in another small cavity between the two inductors. The aluminum oven was thermally insulated and placed in a grounded box to shield the measurement from interference.

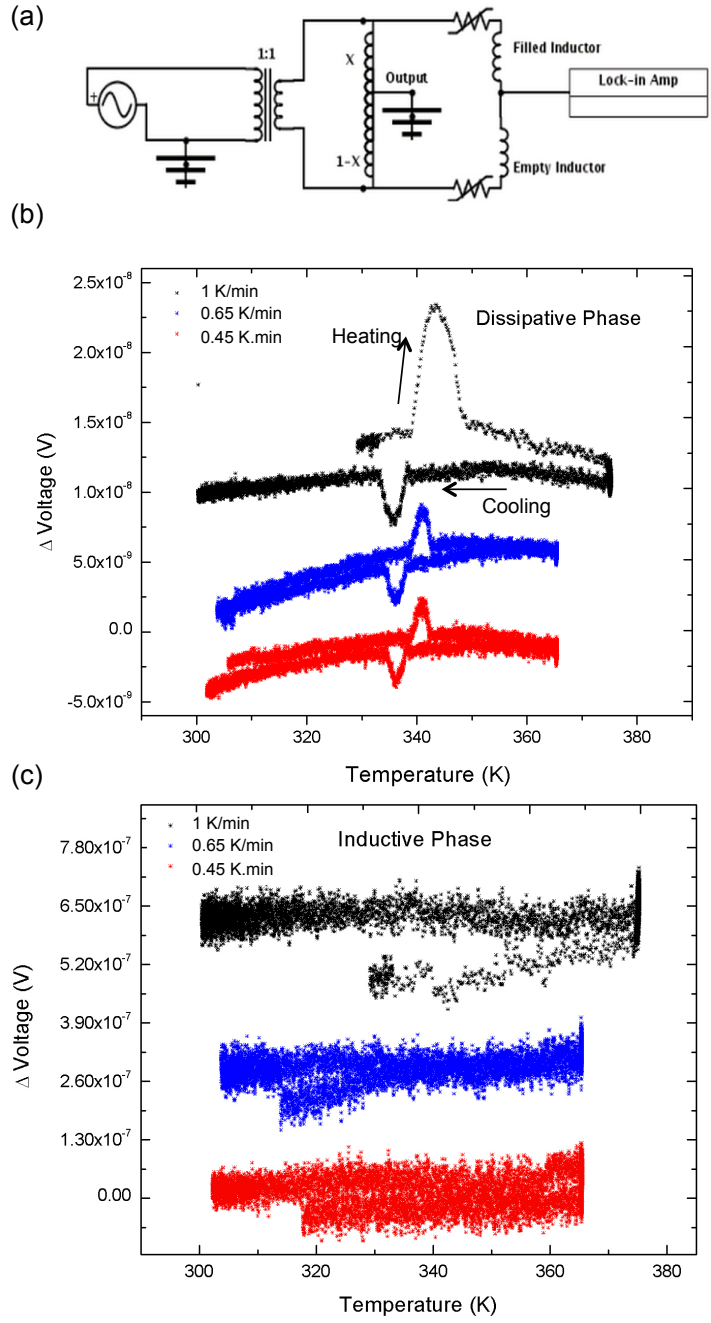

**Figure 2**: (a) AC bridge circuit employed in these measurements. The ratio transformer and variable resistors are used to null both the inductive and dissipative phases of the bridge off-balance signal prior to sweeping the temperature. (b) Off-balance signal vs. temperature for three sweep rates (listed top-to-bottom in the legend) at fixed AC drive of 2.0 V rms at 2.077 kHz. Note that the sweep rate applies to the warming trace, while the cooling rate (approximately 0.5 K/min or below) is set by the coupling of the aluminum block to the ambient environment. (c) Inductive phase of the same data shown in (b). No detectable feature is seen at the transition.

Figure 2b shows the dissipative phase off-balance bridge response as a function of temperature for three characteristic temperature ramping rates, all at 1 K/minute or below, a limit chosen to avoid large temperature differences between the thermometer and the sample. The upper (lower) branch of each trace corresponds to the temperature ramping upward (downward). The heating curves are taken at the various indicated ramping rates, while the cooling curves are slower (approximately 0.5 K/min or below), always limited by the coupling of the aluminum block to the environment; the result is that the dip in the cooling curve is smaller than the peak in the warming curve, particularly for the most rapid warming trace. There is a slight overall trend of the background with temperature due to some small residual difference in the temperature dependence of the responses of the two coils (independent of the $VO_2$ powder), likely the result of slight differences in geometry.

Our simple expectation had been that the off-balance signal would be dominated by the inductive phase and proportional to the change in the inductance of the coil containing the $VO_2$ powder, $\delta V \propto \delta L(T) \propto \delta \chi'(T)$. Instead, the dominant signal detected by the lock-in is at the dissipative phase with respect to the AC driving voltage, and it is clear that the signal more closely resembles $d\chi'/dt = (\partial \chi'/\partial T)(dT/dt)$, with a voltage peak upon warming and a voltage dip upon cooling. Increasing the temperature sweep rate increases the magnitude of the peak (while decreasing the number of data points captured during the sweep). The offset between the peak temperature and the dip temperature is consistent with the thermal hysteresis commonly observed in the MIT of $VO_2$, as in Fig. 1.

There is no readily detectable sign of a change in the inductive response (Fig 2c). This is because the analog of the signal in Fig. 1 is apparently too small to detect. One would expect a step-like function of temperature (and hence time), as in the measured AC susceptibility data shown in Fig. 1. If $f$ is the filling factor of the inductor, total, then the fractional change in inductance should be $f\Delta\chi$, where $\Delta\chi$ is the change in susceptibility across the transition. Assuming a packing density of the $VO_2$ of about 0.6, and a fraction of the inductor volume available to the $VO_2$ of about $(2.25 \text{ mm}/4 \text{ mm})^2$, we find a total filling fraction $f \approx 0.19$. The change in susceptibility is $(7 \times 10^{-6} \text{ emu/g})(4.57 \text{ g/cm}^3)(4\pi) = 4.0 \times 10^{-4}$ in SI units. Thus, the expected fractional change in inductance is $10^{-5}$, near what could be detected in a bridge setup such as this. (The hope in designing experiment was that this would be detectable, but with the noise levels and systematic drifts in the measurement as implemented, this turned out to be overly optimistic.)

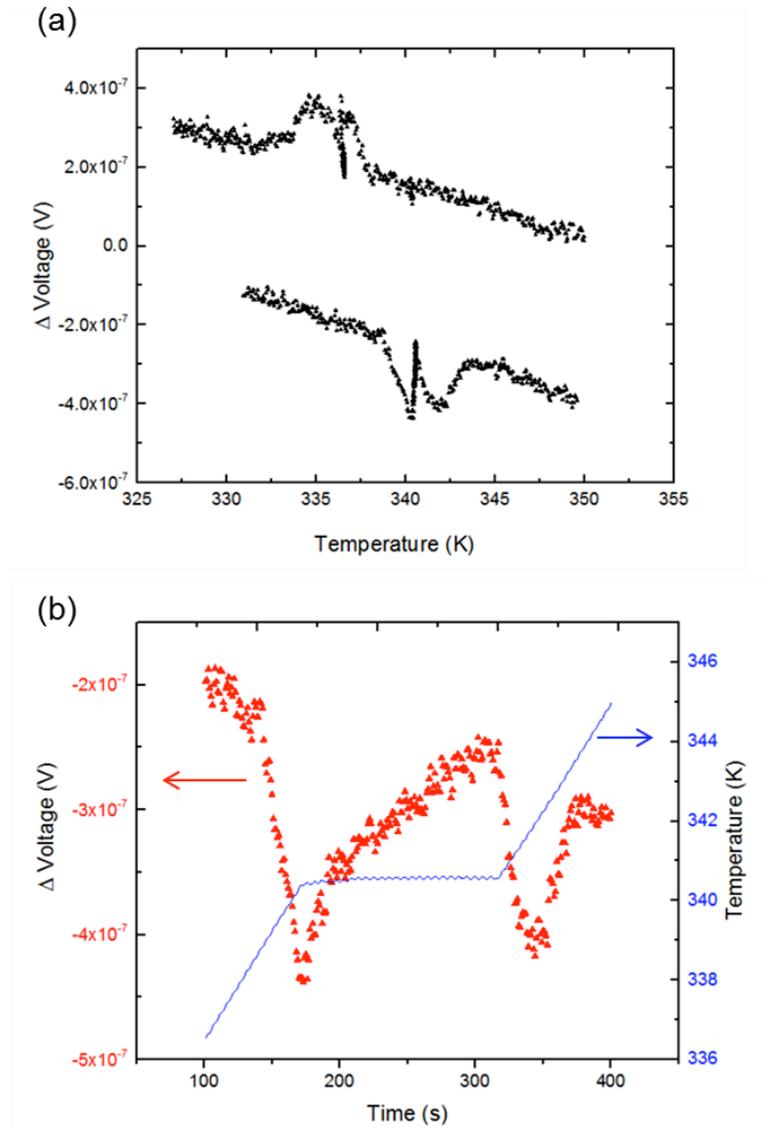

**Figure 3**: (a) Off-balance signal (2 V AC drive, 4.11 kHz) vs. temperature with a pause in the temperature ramp (1.25 K/min) at the middle of each transition   In this figure, the polarity of the bridge (the filled vs. empty inductor positions in the circuit) was deliberately reversed from that used in Fig. 2b, so that a peak is seen upon cooling and vice-versa.  This reversal of sign in the off-balance signal upon rearrangement of the wiring further confirms that the off-balance signal originates with the $VO_2$ powder. (b) Detail of heating transition from (a), with off- balance signal vs. time showing the stop and restart of the temperature sweep.

To confirm the hypothesis that the data were reflecting the effective time derivative of the magnetic susceptibility, we halted the temperature sweep mid-transition.  While the temperature is held fixed, the number of currently transitioning crystals within the sample is expected to approach zero as the sample equilibrates. As shown in Fig. 3a, the off-balance signal relaxes to the value extrapolated from the general warming or cooling trend during these pauses. When the temperature ramp resumes, the peak (or dip) in the off-balance signal re-emerges.

The peak (dip) in the off-balance signal originates from the full inductive response of the filled inductor. In the presence of an AC current, $I = I_0 \exp(-j\omega t)$, the voltage drop across the inductor is expected to be $-d(L(T)I)/dt = j\omega LI - I(dL/dT)(dT/dt)$. In an ordinary inductor this second term in negligible. However, in this particular case, the inductance can be written $L = L_0(1 + f \sum_n \chi_n'(T) m_n)$, where $f$ is a geometric factor associated with the filling of the inductor, $\chi'(T)$ is the temperature dependent susceptibility of a particular grain, and $m_n$ is the mass of that grain. Each grain has its own particular transition temperature upon warming (and cooling), where $\chi'(T)$ is nearly discontinuous. Therefore, even though the susceptibility of a typical grain only changes by about $7 \times 10^{-6}$ emu/g, there can be a contribution to the off-balance voltage that is considerably larger than expected from the resulting change in the classical impedance. In other words, $(dL/dT)(dT/dt) \gg \omega L$ because of the nearly singular contributions of $d\chi'/dT$ from each grain as each grain undergoes the transition. When the temperature ramping is paused, the number of crystals undergoing the transition approaches zero, and the contribution of this term to the off-balance voltage vanishes. We have confirmed using the series resistors in Fig. 1 that the peak (dip) in the off-balance voltage has the appropriate phase relationship with the current as expected from this argument.

If the inductance simply varied smoothly with time, the instantaneous $dL/dt$ would equal the average $dL/dt$. There would be some contribution to the dissipative phase because of that time averaged $dL/dt$. However, there is some (complicated) instrument response function $K(t)$, so that the lock-in output resembles $\Delta V(t) = \int_0^\infty K(t-\tau)(dL/dt(\tau))d\tau$ (The lock-in response can only depend on the current and previous values of $dL/dt$). This response is going to look differently depending on whether or not $dL/dt$ is a smooth function or (because $L(t)$ is piece-wise discontinuous) a succession of nearly singular spikes. Importantly, we believe that if $dL/dt$ did not contain sharp spikes, there would not be a detectable instrument response. Because there is some instrumental noise floor, this implies that the instrument response is technically nonlinear near and below the threshold for detectability.

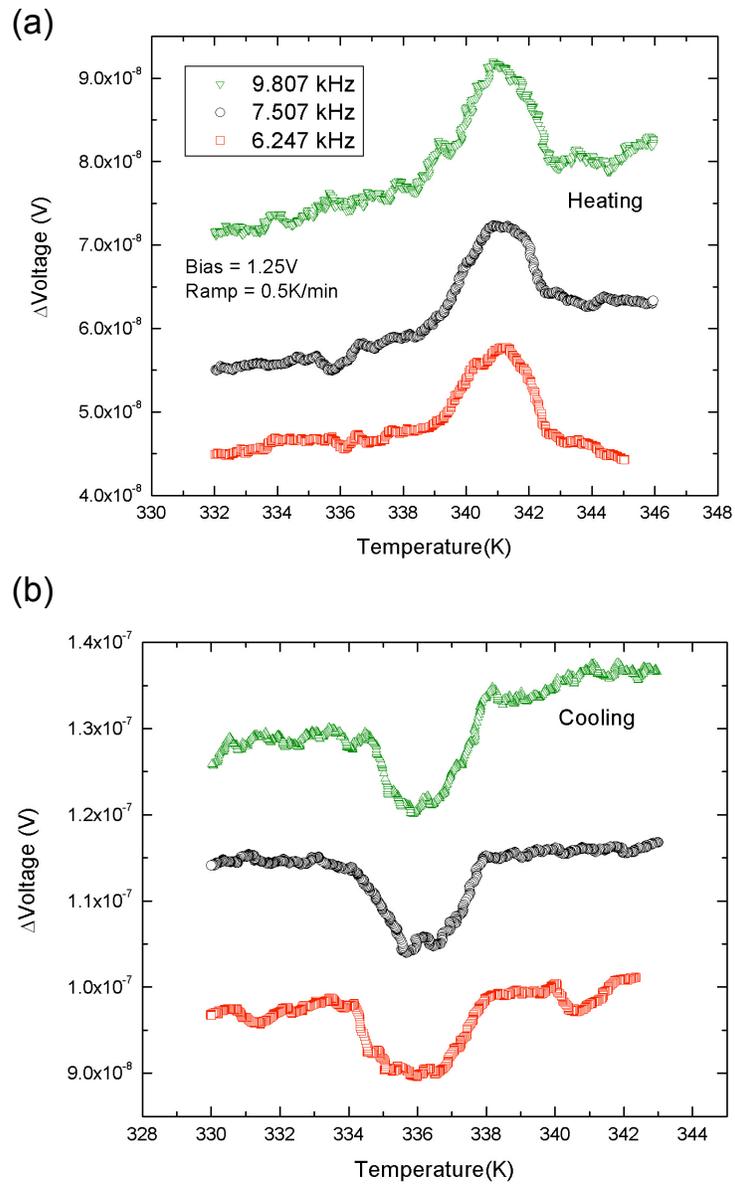

**Figure 4:** The heating peaks (a) and cooling dips (b) in the dissipative response measured at three different frequencies (indicated top-to-bottom in the legend), with equal AC driving voltage and at equal temperature ramping rate.

For quicker ramping rates, each data point represents more crystals transitioning in the period of time captured, thereby increasing the value of $\Delta V(t)$. For smaller ramping rates, each data point represents fewer crystals transitioning, resulting in a lower value. However, there are more points in the curve to compensate. In other words, the time integral of the off-balance voltage response as the entire sample goes through the transition should be roughly the same, provided that the off-balance response is built up from the impulses of the individual grain transitions. Experimentally these time integrals in all measurements upon warming and cooling are identical to within 24%. Given the slight mismatches between warming and cooling curves (see Fig. 2b) due to nonuniformities in the temperature and coils, and the intrinsic timescales associated with the lock-in measurement, this is a reasonable level of agreement.

Fig. 4 shows a detail of the peaks and dips at various frequencies. As can be seen in the plots, any specific difference in the shapes and sizes of the curves is within the noise at our level of precision. In principle there should also be a step across the transition, due to the greater dissipative losses in the metallic phase. However, realistically estimating the magnitude of this effect in a powder is difficult, given the uncertainty in the grain-to-grain coupling. (Commercial inductors frequently use powdered high permeability materials in their cores specifically because the powder morphology strongly reduces eddy currents.) Any dissipative contribution from eddy currents should scale quadratically in bridge excitation amplitude, and quadratically in measurement frequency. Fig. 4 shows that our measured response in contrast is, to within detection limits, essentially independent of frequency over the available. The off-balance signal of the bridge also scales linearly with excitation, with no signature of a step at the transition quadratic in voltage. Therefore, any effect of eddy currents on the dissipative phase is too small to discern in the presence of the signature that we do detect.

The proposed origin for this kind of peak (dip) in response during a temperature sweep is analogous to a thermally driven version of the Barkhausen effect[9]. The ordinary Barkhausen effect occurs in a macroscopic ferromagnet as a function of applied magnetic field. Because of the reorientation of domains and the sudden propagation of domain walls, the smooth averaged magnetization response is actually composed of many nearly discontinuous changes in magnetization as magnetic field is swept. The sudden reconfiguration of domains leads to detectable voltage spikes as a function of time in a pickup coil enclosing the material. Note that Barkhausen noise is not "noise" in the sense of temporal fluctuations away from the average response. It refers specifically to a signal (that sounds like acoustic noise) that is detectable because of sharp, sudden changes in the magnetization. In our experiment, we infer that we, too, have sharp temporal changes in the magnetization. However, in our case these are not occurring as a function of applied magnetic field. Rather, they are taking place as a function of time (and temperature) as the magnetic susceptibility of the material changes suddenly, jumping as grains (or domains) of $VO_2$ suddenly change phase between the monoclinic and rutile states.

The analog to the $VO_2$ experiment occurs when a material passes through a magnetic ordering transition as a function of temperature. Critical fluctuations of the magnetization during a temperature sweep through the transition give an analogous peak (dip) in (slow) measurements of the AC susceptibility, as seen by Ishizuka *et al*. when examining $Mn(HCOO)_2 2H_2O$[10,11]. Analogous effects involving the onset of ferroelectric polarization in ferroelectric materials have also been reported[12,13]. These dynamical effects are not observable in the ACMS data of Fig. 1 because of the standard measurement procedure of waiting to acquire data at each point until the temperature and magnetometer response have stabilized.

Because the lock-in setup uses a kHz-range excitation and a tenth of millisecond time constant for its output, it clearly is not well-suited to direct detection of (MIT-produced, through sudden changes in the magnetic susceptibility as a grain transitions) voltage pulses that could easily be much shorter. However, a higher bandwidth measuring approach could be capable of detecting these spikes, providing experimental access through an electronic measurement to the intrinsic speed of the thermally driven MIT and to the distribution of switching temperatures within such an ensemble of grains.  A very crude estimate of the timescale for the transition to take place in a single grain would be the grain size (on the order of 100 micrometers) divided by the speed of sound (on the order of several km/s), since the transition involves an elastic distortion; the resulting timescale is tens of nanoseconds.  The total inductance is around 10 microHenries.  A reasonable DC current through the inductor would be on the order of 100 mA to avoid self-heating of the coil.  The volume filling fraction of one grain in the inductor would be $1.6 \times 10^{-6}$.  One grain (approximated as a cube 100 microns on a side) would have a mass of about $4.6 \times 10^{-6}$ grams.  Putting this all together, if the transition of the grain takes a time $\Delta t = 30$ ns, the expected voltage pulse should be given by $I \frac{dL}{dt} \approx IL_0 (f \frac{d\chi}{dt}) \approx IL_0 (f \frac{\delta \chi}{\delta t})$.  Plugging in, we would estimate a voltage pulse of $2.1 \times 10^{-8}$ V (and a duration of 30 ns).  This is challenging to measure directly.  However, if instead a static magnetic field of several Tesla were applied to such grains, pulses in the microvolt range (more experimentally accessible) would be possible.

This thermal Barkhausen analogy, which rests on the inadequacy in this situation of the first-year undergraduate physics modeling of an inductor with a simple impedance, should be adaptable to a more general study of first-order (through domain nucleation and growth) and second-order (through critical fluctuations) phase transitions that have some coupling to the magnetic susceptibility.

**Acknowledgments**:

BHR, WJH, HJ, and DN acknowledge support from DOE BES award DE-FG02-06ER46337.